\documentstyle[epsfig]{mn2e}

\title{Intranight optical variability of Blazars}
\author[R. Sagar et al. ]
       {Ram Sagar$^1$, C. S. Stalin$^{1,2,3}$, Gopal-Krishna$^2$ and 
Paul J.\ Wiita$^4$ \\
$^{1}$ State Observatory, Manora Peak, Nainital 263 129, India \\
$^{2}$ National Centre for Radio Astrophysics, TIFR, Pune University Campus, Post Bag 3, Pune 411
007, India\\
$^{3}$ Laboratoire de Physique Corpusculaire et Cosmolgie, College de
France,
11 pl.\ Marcelin Berthelot, F-75231, Paris Cedex 5, France\\
$^{4}$ Department of Physics \& Astronomy, MSC 8R0314, Georgia State University, Atlanta, 
Georgia 30303-3088, USA}

\date{Accepted 2003 October 23, Received 2003 October 3; in original form 2003 May 21}
\pagerange{\pageref{firstpage}--\pageref{lastpage}}
\pubyear{2003}
\begin{document}

\maketitle

\label{firstpage}

\begin{abstract}
We present results of a multi-epoch intra-night optical monitoring of 
eleven blazars consisting of six BL Lac objects and five radio 
core-dominated quasars (CDQs). These densely sampled and sensitive R-band 
CCD observations, carried out from November 1998 through May 2002 during
a total of 47 nights with an average of 6.5 hours/night, have enabled us 
to detect variability amplitudes as low as $\sim 1\%$ on intra-night 
time scales. A distinction is found for the first time between the 
intra-night optical variability (INOV) properties of the these two 
classes of relativistically beamed radio-loud AGNs.  BL Lacs are found 
to show a duty cycle (DC) of INOV of $\sim$60\%, in contrast to 
CDQs, which show a much  smaller INOV DC of $\sim$20\%, the
difference being attributable mainly to the weakly polarized CDQs. 
On longer time scales (i.e., between a week to a few years) variability 
is seen from all the CDQs and BL Lacs in our sample. The results 
reported here form part of our long-term programme to understand the
intra-night optical variability characteristics of the four main 
classes of luminous AGNs, i.e., radio-quiet quasars (RQQs) and radio 
lobe-dominated quasars (LDQs), as well as CDQs and BL Lac objects.

\end{abstract} \begin{keywords}
galaxies: active -- galaxies: jets -- galaxies: photometry -- quasars: general
\end{keywords}

\section{Introduction}

Variability observations of active galactic nuclei (AGNs) on
intra-night time scales can provide valuable clues to 
the physics of the innermost nuclear regions in these
objects. Blazars (core-dominated quasars: CDQs,  and BL Lac objects: BL Lacs) 
as a class of AGNs are characterised by the most violent variations 
at almost all wavelengths over a wide range of time scales.  Blazars' 
properties are consistent with relativistic beaming, that is bulk
relativistic motion of the jet plasma at small angles
to the line of sight, which gives rise to strong amplification
and rapid variability in the observer's frame. CDQs and 
BL Lacs are thought to be the beamed counterparts of 
high and low luminosity radio galaxies, respectively (e.g., Urry 
\& Padovani 1995). The main difference between CDQs and BL Lacs 
lies in their emission lines, which are strong  in CDQs, but weak, 
and in many cases, undetected, in BL Lacs.  CDQ spectra can extend 
up to around GeV energies, whereas the spectra of BL Lacs can extend 
up to TeV energies. Though it is very likely that both CDQs and BL 
Lacs are dominated by non-thermal Doppler boosted jets, some 
important differences have been found between their apparent non-thermal 
properties, such as the magnetic field patterns in their parsec-scale jets 
(Gabuzda et al.\ 1992; but see Gopal-Krishna \& Wiita 1993). 

The intra-night optical variability (INOV), or microvariability, 
of blazars has been an established phenomenon for over a dozen years 
(Miller, Carini \& Goodrich 1989; Carini et al.\ 1991). Although the 
origin(s) of INOV in all AGNs is still uncertain, for blazars it is 
generally associated with the non-thermal Doppler boosted emission from 
jets (Blandford \& Rees 1978; Marscher \& Gear 1985; Marscher, Gear \& 
Travis 1992; Hughes, Aller \& Aller 1992; Wagner \& Witzel 1995). 
Still,  alternative models, which invoke accretion disk instabilities
or perturbations (e.g., Mangalam \& Wiita 1993; for a review, see Wiita 1996) may
also explain some INOV, particularly in radio-quiet quasars (RQQs) where any 
contribution from the jets, if they are at all present, is weak. Several studies 
of the INOV of blazars are available in the literature (e.g., Heidt \& Wagner 1996; 
 Dai et al.\ 2001; Romero et al.\ 2002; Xie et al.\ 2002). However, the unique
feature of the present study is the deliberate focus on the comparison of the 
INOV properties of the two Doppler beamed AGN classes, namely CDQs and BL Lacs. 
Other results from this large programme, involving the nature of INOV
in BL Lacs and RQQs (Gopal-Krishna et al.\ 2003, hereafter GSSW03), and a 
comparison of INOV between lobe-dominated radio-loud quasars (LDQs) and 
RQQs (Stalin et al.\ 2003a),  have been published elsewhere.

\begin{table*}
\caption{The sample of core-dominated quasars and BL Lacs monitored in the present programme}
\begin{tabular}{clllllllrrr} \hline \hline
Object       & Other Name & Type & RA(2000)  & Dec(2000)   &  ~~B    & ~~M$_B$  & {\it ~~z} & $\alpha$~  & \%Pol$^\ast$ &
R$^{\dag}~$\\
             &            &      &           &             & (mag)   & (mag)    &           &           & (opt)       &  \\ \hline
0219+428   & 3C 66A         & BL  & 02 22 39.6 &   +43 02 08 & 15.71  & $-$26.5 & 0.444   & $-$0.19$\S$ &11.70 &  676.1 \\
0235+164   & AO 0235+164    & BL  & 02 38 38.9 &   +16 37 00 & 16.46  & $-$27.6 & 0.940   &    0.67~      &14.90 & 1949.8 \\
0735+178   & PKS 0735+17    & BL  & 07 38 07.4 &   +17 42 19 & 16.76  & $-$25.4 &$>$0.424 & $-$0.26$\S$ &14.10 & 3548.1 \\
0851+202   & OJ 287         & BL  & 08 54 48.8 &   +20 06 30 & 15.91  & $-$25.5 & 0.306   &    0.18$\S$ &12.50 & 2089.3 \\ 
0955+326   & 3C 232         & CDQ & 09 58 20.9 &   +32 24 02 & 15.88  & $-$26.7 & 0.530   & $-$0.11$\S$ & 0.53 &  549.5 \\
1128+315   & B2 1128+31     & CDQ & 11 31 09.4 &   +31 14 07 & 16.00  & $-$25.3 & 0.289   & $-$0.41~      & 0.62 &  269.2 \\   
1215+303   & B2 1215+30     & BL  & 12 17 52.0 &   +30 07 01 & 16.07  & $-$24.8 & 0.237   & $-$0.17$\S$  & 8.00 &  426.6 \\ 
1216$-$010 & PKS 1216$-$010 & CDQ & 12 18 35.0 & $-$01 19 54 & 16.17  & $-$25.9 & 0.415   & $-$0.03$\S$ & 6.90 &  218.8 \\
1225+317   & B2 1225+31     & CDQ & 12 28 24.8 &   +31 28 38 & 16.15  & $-$30.0 & 2.219   &    0.01~      & 0.16 &  182.0 \\ 
1308+326   & B2 1308+32     & BL  & 13 10 28.7 &   +32 20 44 & 15.61  & $-$28.6 & 0.997   & $-$0.09$\S$ &10.20 &  512.9 \\
1309+355   & PG 1309+355    & CDQ & 13 12 17.7 &   +35 15 23 & 15.60  & $-$24.7 & 0.184   & $-$0.12~      & 0.31 &   22.9 \\ \hline
\end{tabular}

\hspace*{-10.3cm} $^\ast$ Reference for optical polarizations: Wills et al.\ (1992)

\hspace*{-7.8cm} $^\dag$ R is the ratio of the radio-to-optical flux densities as defined in the text

\hspace*{-7.7cm} $\S$ Reference for these radio fluxes: Kovalev et al.\ (1999); for others see text

\end{table*}

\section[]{Sample, observations and Reductions}

The sample of blazars  used in this work consists of 6 BL Lac objects and 
5 CDQs.  All are bright, with apparent $B$ magnitudes between 15.6 and 16.8,
so that short exposures can still provide good signal to noise ratios. At most 
two of these, one CDQ (1225$+$317, with $z$ = 2.219) and possibly one BL Lac 
(0735$+$178, with $z > 0.424$), lie at $z > 1$, so this sample provides a fairly 
even coverage of the redshift range up to $z$ = 1 for each blazar subclass. Note 
that we have adopted a BL Lac classification for B2 1308$+$326, following the 1-Jy 
(Stickel, Fried \& K{\"u}hr 1993) and Padovani \& Giommi (1995) catalogs, even 
though it is classified as a CDQ in the V{\'e}ron-Cetty \& V{\'e}ron (1998) catalog. 
Also, for the BL Lac object PKS 0735$+$178 we have adopted a redshift $z$ = 0.424, though
formally the published value is $z > 0.424$ (see V\'eron-Cetty \& V\'eron 1998). For each object we
could find a measurement of the degree of optical polarization (Wills et al.\ 1992); it
can be seen from Table 1 that for four out of the five CDQs the percentage
polarization was $<$ 1\% at the time those measurements were made, and 
hence they can be assigned to the sub-class of low-polarization CDQs.

Basic information of these objects are given in Table 1. The values of the radio 
spectral index, $\alpha$ (with $S_{\nu} \propto \nu^{\alpha}$), given in Table 1 
were usually determined from linear spectral fitting to the available {\it near 
simultaneous} flux density measurements between 1 and 22 GHz reported by 
Kovalev et al. (1999); those not appended with $\S$ are based on 
linear fits to {\it non-simultaneous} flux measurements taken 
from NED\footnote{URL http://ned.ipac.caltech.edu/}. 

The observations were made using the 104-cm Sampurnanand telescope of
the State Observatory, Naini Tal which is an RC system with a f/13 beam 
(Sagar 1999). The detectors used were a cryogenically cooled  1024 $\times$ 1024
CCD chip (prior to October 1999) and  a 2048 $\times$ 2048 chip (after October 
1999), both mounted at the Cassegrain focus. Each pixel of both the CCDs correspond 
to a square of 0.38 arcsec on the sky, covering a total field of $\sim$
12$^\prime$ $\times$ 12$^\prime$ in the case of the larger CCD and $\sim$ 
6$^\prime$ $\times$ 6$^\prime$ in the case of the smaller CCD. Observations 
were almost always done using an R filter, as it was near the maximum response of
the CCD system and thus allowed us to achieve good temporal resolution;
however, on two nights quasi-simultaneous observations were done using R and I 
filters.  To improve S/N, observations were carried out in 2 $\times$ 2 
binned mode. On each night only one QSO was monitored as continuously as possible
and the typical sampling rate was about 5 frames per hour. The choice
of the exposure time depended on the brightness state of the QSO, the 
moon's phase and sky transparency. The field containing the QSO was
adjusted so as to have at least 2 (and usually 3) comparison stars within about
a magnitude of the QSO on the CCD frame. 

Preliminary processing of the images as well as photometry was done using the 
IRAF\footnote{Image Reduction and Analysis Facility, distributed by NOAO, operated 
by AURA, Inc. under agreement with the US NSF.} software. Photometry of the 
QSO and the comparison stars recorded on the same CCD frame was carried out 
using the {\it phot} task in IRAF. The same circular aperture was used for the 
photometry of the QSO and the comparison stars for all the images acquired over 
the night. This optimum aperture was selected by considering a range of apertures 
starting from the median FWHM over the night for the photometry and choosing that 
aperture that produced the minimum variance in the star -- star 
differential lightcurve (DLC) of the steadiest pair
of comparison stars. Further details of the observations and reductions
are presented elsewhere (Stalin 2002; Stalin et al.\ 2003b). DLCs of the 
AGN relative to the comparison stars as well as between all pairs of 
comparison stars (usually three, but in some cases, two)
are constructed from the derived instrumental magnitudes. 
The DLCs of the AGN relative to the comparison stars
are used to look for the presence of 
INOV in the AGN. The choice of more than one comparison star in the 
differential photometry enables us to reliably identify QSO variability,
as any stars which themselves varied during the night can be identified and discarded. The
position and apparent magnitudes of the comparison stars used in the  
differential photometry of our sample of blazars from the USNO 
catalog\footnote{http://archive.eso.org/skycat/servers/usnoa} are given in Table 2.
Note that the magnitudes of the comparison stars taken from this
catalog have uncertainties of up to 0.25-mag.
\begin{figure*}
\vspace*{-0.8cm}
\hspace*{-1.0cm}\psfig{file=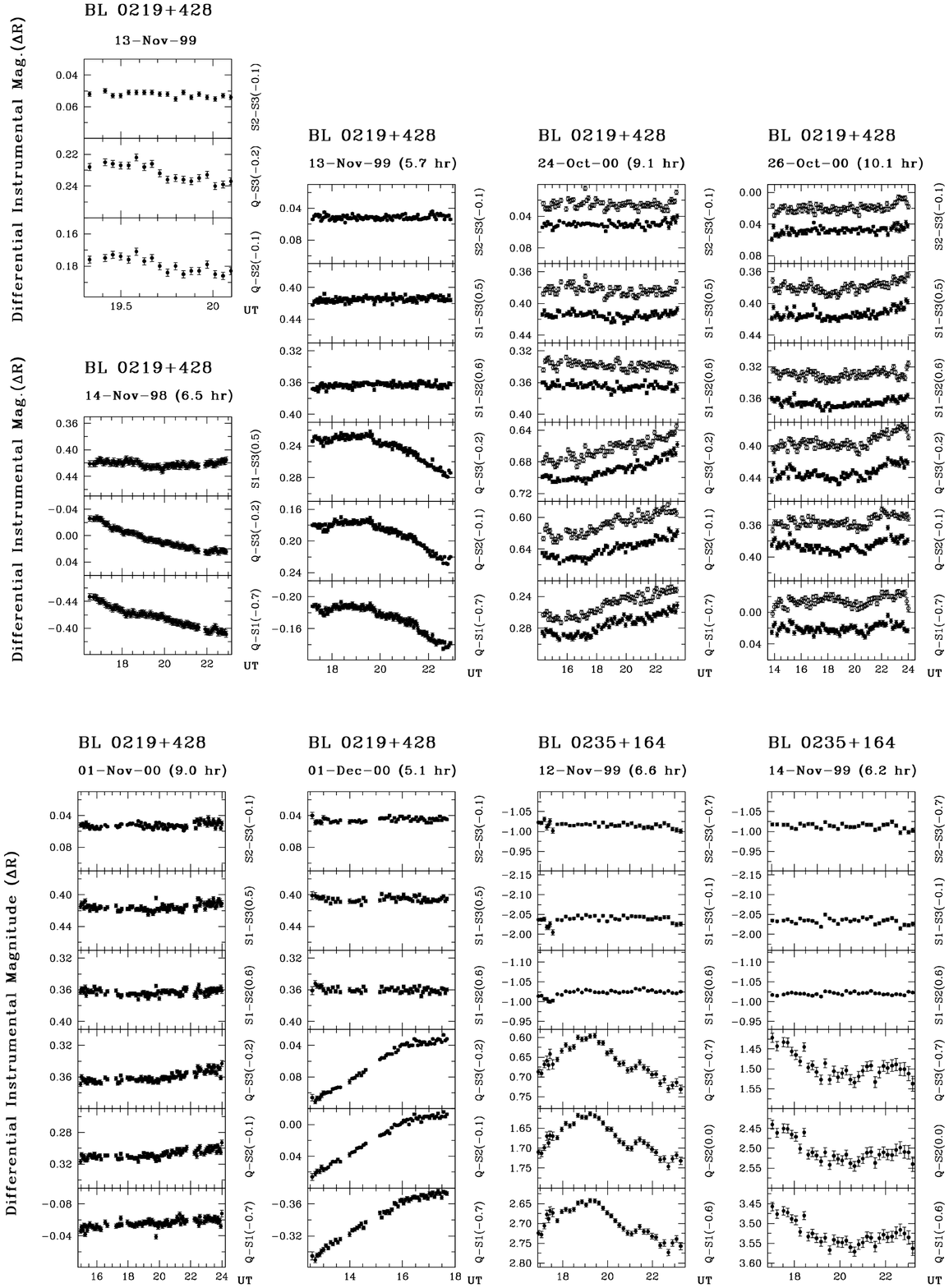,width=18cm,height=23cm}
\noindent{\bf Figure 1.} Differential R-band light curves for the BL Lacs and core-dominated
quasars. The name of the object, the date and duration of observation are given on the top of each panel. The upper
panel(s) give the DLC of the pair(s) of comparison stars whereas the subsequent lower panels
are the DLCs of the quasar relative to the comparison stars. Note the different scales
for many of the sources.  I-band DLCs are shown as open symbols for two nights for BL 0219$+$428.
\end{figure*}
\begin{figure*}
\hspace*{-1.0cm}\psfig{file=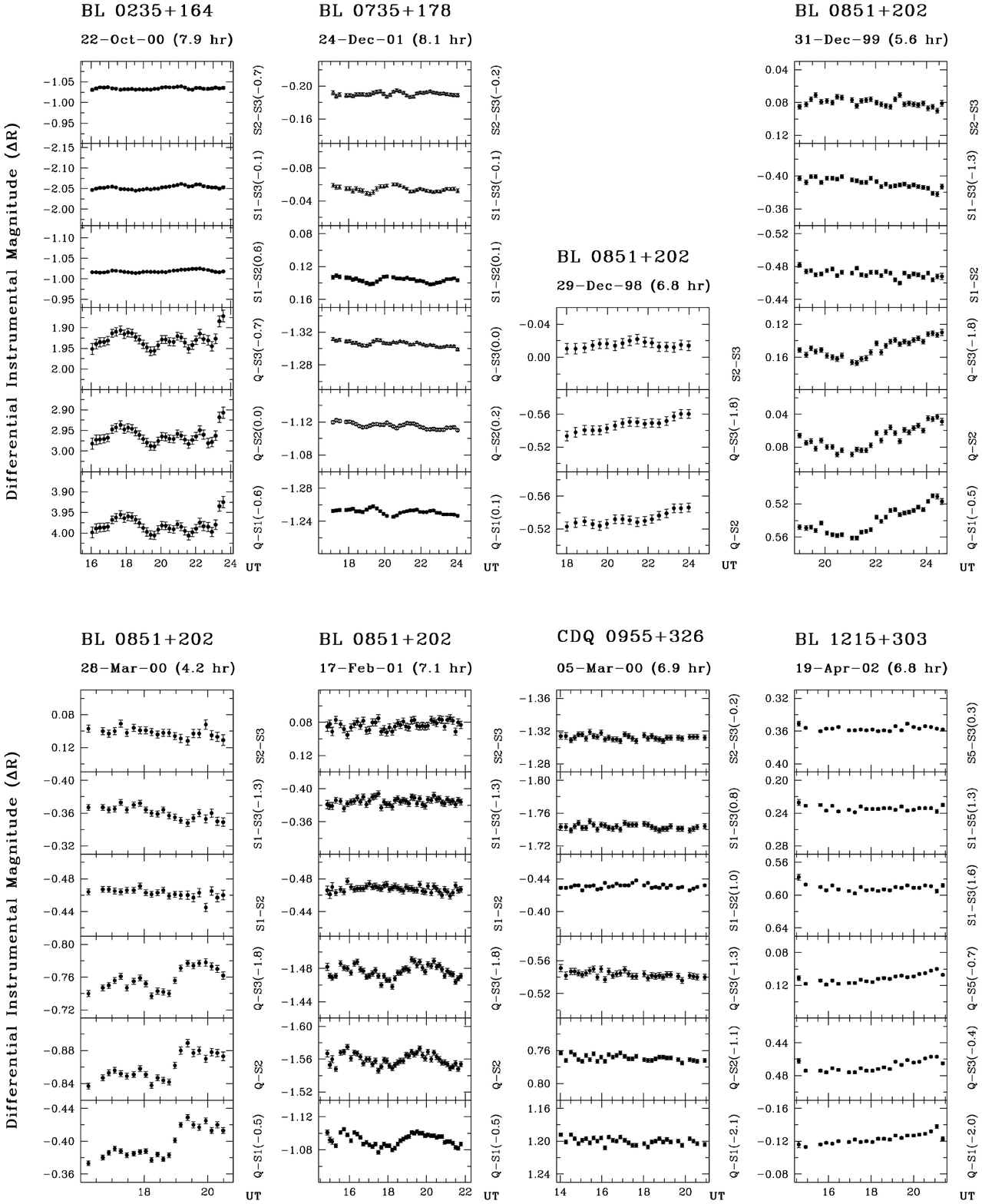,height=23cm,width=18cm}
\noindent{\bf Figure 1.} {\it Continued}
\end{figure*}

\begin{figure}
\hspace*{1.0cm}\psfig{file=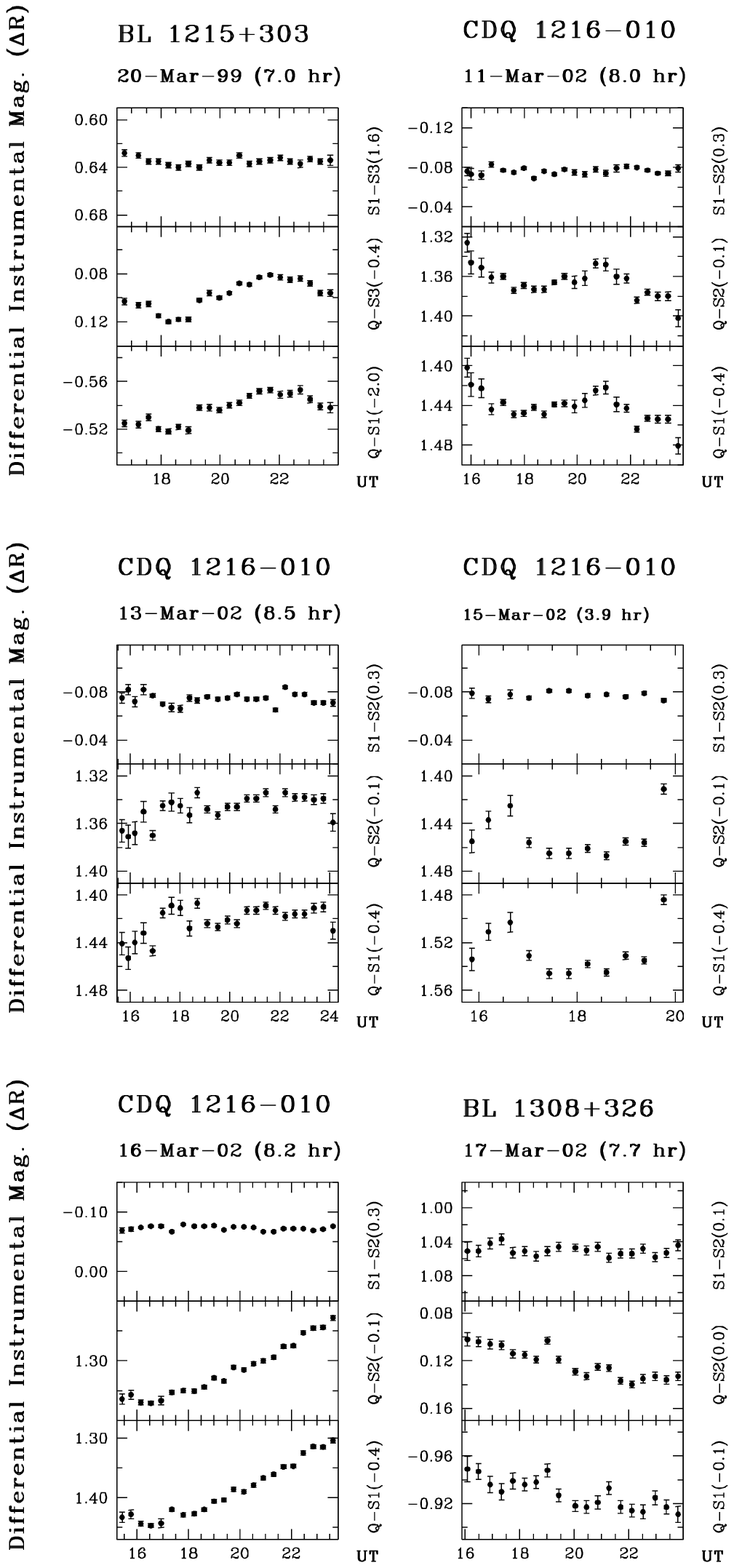,width=14cm, height=18cm}
\hspace*{1.5cm}{\bf Figure 1.} {\it Continued}
\end{figure}

\section{Results}
\subsection{Differential light curves (DLCs)} DLCs are presented for 
those AGNs which have shown clear evidence of INOV in Fig.\ 1. 
We consider a source to be variable only if correlated variations 
(both in amplitude and time) are found in the DLCs of the AGN relative 
to all the comparison stars considered.  All of the six BL Lacs in our 
sample showed INOV on at least one night, whereas this was the case for
only two of the five CDQs monitored (Table 3).  For each variable AGN, 
we have statistically quantified the variability and have derived 
variability parameter, amplitude and time scale of variability below.

\begin{table}
\caption{Positions and apparent magnitudes of the comparison stars}
\begin{tabular}{lcllcc} \hline
 Source     &   Star    &     RA(2000)   & Dec(2000)    & R      &   B         \\
            &           &    h ~ m ~ s     &  d ~ m ~ s     & mag    &  mag        \\ \hline
 0219+428   &   S1      &   02 22 45.13  &  43 04 19.6  &  14.2  &  15.6       \\
 BL         &   S2      &   02 22 47.23  &  43 06 00.1  &  13.9  &  14.7       \\
            &   S3      &   02 22 28.39  &  43 03 40.7  &  13.9  &  14.8       \\
 0235+164   &   S1      &   02 38 56.44  &  16 38 56.5  &  14.5  &  15.7       \\
 BL         &   S2      &   02 38 54.48  &  16 36 03.1  &  15.5  &  16.1       \\
            &   S3      &   02 38 38.53  &  16 40 05.2  &  16.7  &  18.0       \\
 0735+178   &   S1      &   07 38 03.45  &  17 42 56.1  &  16.4  &  16.5       \\
 BL         &   S2      &   07 38 17.10  &  17 39 03.7  &  16.0  &  16.0       \\
            &   S3      &   07 38 10.29  &  17 43 43.9  &  16.4  &  16.6       \\
 0851+202   &   S1      &   08 54 46.11  &  20 07 20.3  &  15.7  &  16.8       \\
 BL         &   S3      &   08 54 43.70  &  20 02 42.2  &  16.4  &  18.8       \\
 0955+326   &   S1      &   09 58 14.45  &  32 23 45.8  &  14.0  &  15.0       \\
 CDQ        &   S2      &   09 58 18.32  &  32 28 35.1  &  14.5  &  14.9       \\
            &   S3      &   09 58 26.47  &  32 26 54.3  &  15.9  &  16.5       \\
 1128+310   &   S1      &   11 31 02.12  &  31 11 39.3  &  15.9  &  16.7       \\
 CDQ        &   S2      &   11 30 54.41  &  31 11 47.8  &  15.6  &  15.8       \\
            &   S3      &   11 31 18.04  &  31 17 16.8  &  15.5  &  17.1       \\
 1215+303   &   S1      &   12 17 45.96  &  30 04 51.0  &  14.4  &  16.8       \\
 BL         &   S2      &   12 17 49.13  &  30 07 02.3  &  15.8  &  16.4       \\
            &   S3      &   12 17 44.47  &  30 09 44.1  &  13.7  &  14.5       \\
            &   S4      &   12 18 09.03  &  30 09 35.8  &  14.7  &  15.6       \\
            &   S5      &   12 17 26.62  &  30 07 53.5  &  14.2  &  15.3       \\
 1216-010   &   S1      &   12 18 42.91  & -01 19 24.8  &  15.4  &  16.2       \\
 CDQ        &   S2      &   12 18 45.07  & -01 19 47.2  &  15.3  &  15.8       \\
 1225+317   &   S1      &   12 28 18.78  &  31 25 20.1  &  14.6  &  15.5       \\
 CDQ        &   S2      &   12 28 30.62  &  31 26 34.2  &  15.6  &  16.6       \\
            &   S3      &   12 28 13.60  &  31 27 36.3  &  17.1  &  18.4       \\
            &   S4      &   12 28 29.15  &  31 25 18.5  &  15.5  &  16.7       \\
 1308+326   &   S1      &   13 10 19.69  &  32 23 53.6  &  16.9  &  17.8       \\
 BL         &   S2      &   13 10 18.09  &  32 20 07.3  &  15.8  &  16.6       \\
            &   S3      &   13 10 29.69  &  32 25 54.0  &  16.4  &  16.8       \\
            &   S4      &   13 10 38.31  &  32 17 31.7  &  16.0  &  16.7       \\
 1309+355   &   S1      &   13 12 39.17  &  35 19 50.4  &  14.6  &  16.0       \\
CDQ         &   S2      &   13 12 40.49  &  35 16 03.0  &  15.0  &  15.9       \\
            &   S3      &   13 12 30.99  &  35 16 08.7  &  14.3  &  15.8       \\ \hline
\end{tabular}
\end{table}

\subsubsection{Variability parameter ($C_{\rm eff}$)}
To quantify the variability, we have employed a statistical criterion
based on the parameter $C$, similar to that
followed by Jang \& Miller (1997), with the added advantage that for each
AGN we have DLCs relative to multiple comparison stars. This allows us to
discard any variability candidates for which the multiple DLCs do not show
clearly correlated trends, both in amplitude and time. We define $C$ for a
given DLC as the ratio of its standard deviation, $\sigma_T$, and the
mean $\sigma$ of its individual data points, $\eta\sigma_{\rm err}$.
Here $\eta$ is the factor by which the average of the measurement 
errors ($\sigma_{\rm err}$, as given by {\it phot}) should be 
multiplied. It has been argued in the literature that the final 
errors given by DAOPHOT/IRAF are often too small (Gopal-Krishna et al.\ 1995; 
Garcia et al.\ 1999). We find $\eta$ = 1.50 (Stalin 2002; GSSW03).
The value of $C_i$ for the $i^{th}$ DLC of the AGN has the corresponding
probability, $p_i$, that the DLC is
steady (non-variable), assuming a normal distribution. For a given AGN
we then compute the joint probability, $P$, by multiplying the values of
$p_i$'s for individual DLCs available for the AGN. The effective
$C$ parameter, $C_{\rm eff}$, corresponding to $P$, is given in Table 3 for
each variable AGN (i.e., $C_{\rm eff} > 2.57$, corresponds to a confidence
level of variability in excess of 99\%).  
We also note that for these AGN all the DLCs involving only 
comparison stars were found to show statistically insignificant
variability. This is quantified by giving within parentheses the values
of $C_{\rm eff}$ for the DLC involving two stable comparison stars. 

\subsubsection{Amplitude of variability ($\psi$)}
For objects which are variable we define the variability amplitude as
(Romero, Cellone \& Combi 1999)
\begin{equation}
\psi = \sqrt{(D_{max} - D_{min})^2 - 2\sigma^2},
\label{amplitude}
\end{equation}
\noindent with
\begin{tabbing}
11111   \=   111111111111111111111111111111111111111111111111111111 \= \kill
$D_{max}$  \> = ~~~~maximum in the quasar differential light curve       \\
$D_{min}$  \>= ~~~~minimum in the quasar differential light curve       \\
$\sigma^2$   \>= ~~~~$\eta^2 \langle \sigma_{err}^2\rangle$. \\
\end{tabbing}

The variability amplitudes computed using Eq.\ (1) for objects
which have shown INOV are given in Table 3 in per cent. Note that the
smallest clearly detected amplitude of INOV in the present sample is 1\%.

\subsubsection{Structure function}

The structure function is frequently used to characterize variability properties
such as time-scales and periodicities present in  light curves. The first order
structure function for a DLC containing N evenly spaced data points is defined as
(Simonetti, Cordes \& Heeschen 1985)

\begin{equation}
D_X^1(\tau) = \frac{1}{N(\tau)}\sum_{i=1}^{N}w(i)w(i + \tau)[X(i + \tau) - X(i)]^2
\end{equation}

\noindent where $\tau$ = time lag, $N(\tau)$ = $\sum w(i)w(i + \tau)$ and the 
weighting factor $w(i)$ is 1 if a measurement exists for the $i^{th}$ interval, 0 
otherwise. The error in each point in the computed structure function is

\begin{equation}
\sigma^2(\tau) = \frac{8\sigma^2_{\delta X}}{N(\tau)} D_X^1(\tau)
\end{equation}
\noindent where $\sigma^2_{\delta X}$ is the measurement noise variance.

Since the samplings of our DLCs are quasi-uniform, we have determined
structure functions using an interpolation algorithm. For any time
lag $\tau$, the value of $X(i+\tau)$ was calculated by linear interpolation
between the two adjacent data points. A typical time scale in the light
curve (i.e., the time between a maximum and  a minimum, or vice versa) is
indicated by a local maximum in the structure function. In case of a monotonically
increasing structure function, the source possesses no typical time-scale
shorter than  the total duration of observations. The plots of the structure
function for objects which have shown definite micro-variability are given
in Fig.\ 2, and the inferred time scales of variabilities are given in Table 3. The
behaviour exhibited by these SF plots for our observations are the following types:
\begin{enumerate}
\item Sources whose SF display no plateau even at long time lag.
The interpretation is that any characteristic time-scale, $\tau$, is longer than
the duration of the light curve and therefore the longest time lag
available represents only  a lower limit to any characteristic time-scale of variation;
\item Sources whose SF show two plateaux.  This indicates the presence of
two time-scales, which may possibly be related to different physical processes.
\item Sources which exhibit one plateau followed by a dip in the structure
function.  The plateau is interpreted as the variability time-scale and the
dip as a period of a possibly cyclic signal in the light curve; these
possible periods are denoted by $P$ in Table 3.
\end{enumerate}

\begin{table*}
\caption{Log of INOV and LTOV observations of CDQs and BL Lacs. In the case of INOT, the variability 
parameter ($C_{\rm eff}$), amplitude ($\psi$), characteristic time scale ($\tau$) and Period (P) of variability
are also given. The number within parentheses in the $C_{\rm eff}$ column denotes the value for the
DLC involving two stable comparison stars. }
\begin{tabular}{lllccrrrrr} \hline
Object     & Date     & No. of & Duration    & INOV  & $C_{\rm eff}$& $\psi$~~  &  $\tau$ ~~ & P ~~ &LTOV\\
    &    & points & (hours)     & status$^*$&AGN(star)          &  (\%)       & (hours) & (hours)&($\Delta$ m)   \\ \hline
0219+428     & 14.11.98 &  118   & 6.5         &  V    & 6.0 (0.5)    & 5.4     &  $>$ 6.5 &      &         \\
  BL           & 13.11.99 &  123   & 5.7         &  V    & $>$ 6.6(1.2) & 5.5     &  $>$ 5.9 &      & $+$0.25 \\
             & 24.10.00 &  73    & 9.1         &  V    & 5.8 (1.5)    & 4.3     &  $>$ 9.1 &      & $+$0.45 \\
             & 26.10.00 &  82    &10.1         &  V    & 3.5 (1.4)    & 3.2     &      4.9 &      & $-$0.26 \\
             & 01.11.00 &  103   & 9.0         &  V    & 2.9 (1.1)    & 2.2     &      3.9 &      & $-$0.08 \\
             & 24.11.00 &  71    & 5.1         &  NV   &              &         &          &      & $+$0.06 \\
             & 01.12.00 &  59    & 5.1         &  V    & $>$6.6 (1.1) & 8.0     &  $>$ 5.1 &      & $-$0.35 \\
0235+164     & 13.11.98 &  36    & 4.4         & ---   &  --------    & ----    & -----    & ---- &         \\
 BL            & 12.11.99 &  39    & 6.6         &  V    & $>$6.6 (1.1) & 12.8    &      3.6 &      & $-$0.67 \\
             & 14.11.99 &  34    & 6.2         &  V    &  3.2 (1.1)   & 10.3    &      3.4 &      & $+$0.83 \\
             & 22.10.00 &  39    & 7.9         &  V    &  2.6 (1.4)   & 7.6     &          &      & $+$0.45 \\
             & 28.10.00 &  29    & 6.8         & ---   & --------     & ----    & -----    & ---- & $+$0.35 \\ 
0735+178     & 26.12.98 &  49    & 7.8         &  NV   &              &         &          &      &         \\
  BL           & 30.12.99 &  65    & 7.4         &  NV   &              &         &          &      & $-$0.45 \\
             & 25.12.00 &  43    & 6.0         &  NV   &              &         &          &      & $-$0.70 \\
             & 24.12.01 &  43    & 7.3         &  V    & 2.8 (2.0)    & 1.0     &  $>$ 8.1 &      & $+$0.43 \\ 
0851+202     & 29.12.98 &  19    & 6.8         &  V    & 2.8 (0.4)    & 2.3     &  $>$ 6.8 &      &         \\
   BL          & 31.12.99 &  29    & 5.6         &  V    & 6.5 (1.1)    & 3.8     &      3.0 &      & $+$0.60 \\
             & 28.03.00 &  22    & 4.2         &  V    & 5.8 (0.8)    & 5.0     &      1.2 &      & $-$0.93 \\
             & 17.02.01 &  48    & 6.9         &  V    & 2.7 (0.7)    & 2.8     &      2.0 & 3.8  & $-$0.70 \\
0955+326    & 19.02.99 &  36    & 6.5         &  NV   &              &         &          &      &         \\
 CDQ            & 03.03.00 &  37    & 6.3         &  NV   &              &         &          &      & $-$0.05 \\
             & 05.03.00 &  34    & 6.9         &  PV   & 2.2 (0.6)    & 0.7     &          &      &    0.00 \\
1128+315    & 18.01.01 &  31    & 5.7         &  NV   &              &         &          &      &         \\
 CDQ           & 09.03.02 &  27    & 8.2         &  NV   &              &         &          &      & $-$0.14 \\
            & 10.03.02 &  28    & 8.3         &  NV   &              &         &          &      &    0.00 \\
1215+303    & 20.03.99 &  21    & 7.0         &  V    & 5.5 (0.9)    & 3.5     &      4.2 &      &         \\
  BL          & 25.02.00 &  28    & 5.9         &  NV   &              &         &          &      & $+$0.19 \\
            & 31.03.00 &  27    & 5.0         &  NV   &              &         &          &      & $+$0.13 \\
            & 25.04.01 &  29    & 6.5         & ---   & --------     & ----    &  -----   & ---- & $+$0.21 \\ 
            & 19.04.02 &  23    & 6.8         &  V    & 4.9 (1.3)    & 1.8     &  $>$ 6.8 &      & $-$0.11 \\
1216$-$010  & 11.03.02 &  22    & 8.0         &  V    & 3.2 (0.9)    & 7.3     &  1.8     & 3.2  &         \\
  CDQ          & 13.03.02 &  24    & 8.5         &  V    & 2.6 (1.5)    & 3.8     &  1.2     & 2.2  & $-$0.02 \\ 
            & 15.03.02 &  11    & 3.9         &  V    & 3.9 (0.9)    & 5.5     &  1.0     & 2.2  & $+$0.11 \\ 
            & 16.03.02 &  22    & 8.2         &  V    & 6.6 (1.4)    & 14.1    &  $>$ 8.2 &      & $-$0.14 \\ 
1225+317    & 07.03.99 &  49    & 6.6         &  NV   &              &         &          &      &         \\
CDQ            & 07.04.00 &  23    & 6.0         &  NV   &              &         &          &      &    0.00 \\
            & 20.04.01 &  34    & 7.4         &  NV   &              &         &          &      & $-$0.02 \\ 
1308+326    & 23.03.99 &  17    & 6.0         & ---   & --------     & ----    & -----    & ---- &         \\
BL            & 26.04.00 &  16    & 5.6         &  NV   &              &         &          &      & $+$0.21 \\
            & 03.05.00 &  19    & 6.7         & ---   & --------     & ----    & -----    & ---- &    0.00 \\ 
            & 17.03.02 &  19    & 7.7         &  V    & 3.1 (0.6)    & 3.4     & 1.2,4.4  &      & $-$1.90 \\ 
            & 20.04.02 &  14    & 5.8         &  NV   &              &         &          &      & $+$0.44 \\ 
            & 02.05.02 &  15    & 5.1         &  NV   &              &         &          &      & $-$0.03 \\ 
1309+355    & 25.03.99 &  39    & 6.7         &  NV   &              &         &          &      &         \\
CDQ            & 01.04.01 &  32    & 4.6         &  NV   &              &         &          &      & $+$0.10 \\
            & 02.04.01 &  41    & 5.2         &  NV   &              &         &          &      &    0.00 \\ \hline
\end{tabular}

\hspace*{-8.0cm} $*$ V = variable, NV = not variable, PV = probably variable

\end{table*}

\subsection{Long term optical variability (LTOV)}
Our observations also provide information on the LTOV. The 
number of epochs covered range between three and seven. For 
the six BL Lacs in our sample the total time-spans covered range 
between about two to three years. Similarly, year-like time coverage 
is available for four of our five CDQs; for the CDQ 1216$-$010 
the overall time-spans are much shorter ($\sim$ 1 week).
Interestingly, even in this case, LTOV is convincingly detected 
just as in the case of the remaining CDQs and BL Lacs in our sample. 
The smallest amplitude of LTOV was found in the CDQ 1225+317; a 
$\sim$ 2\% brightening within a year between April 2000 and April 2001. 
It is noteworthy that this CDQ has not only the highest redshift 
({\it z} = 2.219) but also the lowest optical polarization ($P_{opt}$ = 
0.16\%) in our sample.  Also a $\sim$ 2\% brightening is noticed in 
1216$-$010 within 48 hours.  

The LTOV results are summarized in the last column of Table 3, where 
the differences in the nightly means from those of the previous 
observations are tabulated. Consequently, for an object, the last 
column is left blank at the first epoch of its observations. 
We also give comments on the LTOV of individual sources below.

\section{Notes on the variability of individual sources}

\begin{figure*}
\vspace*{-0.5cm}
\hspace*{-1.0cm}\psfig{file=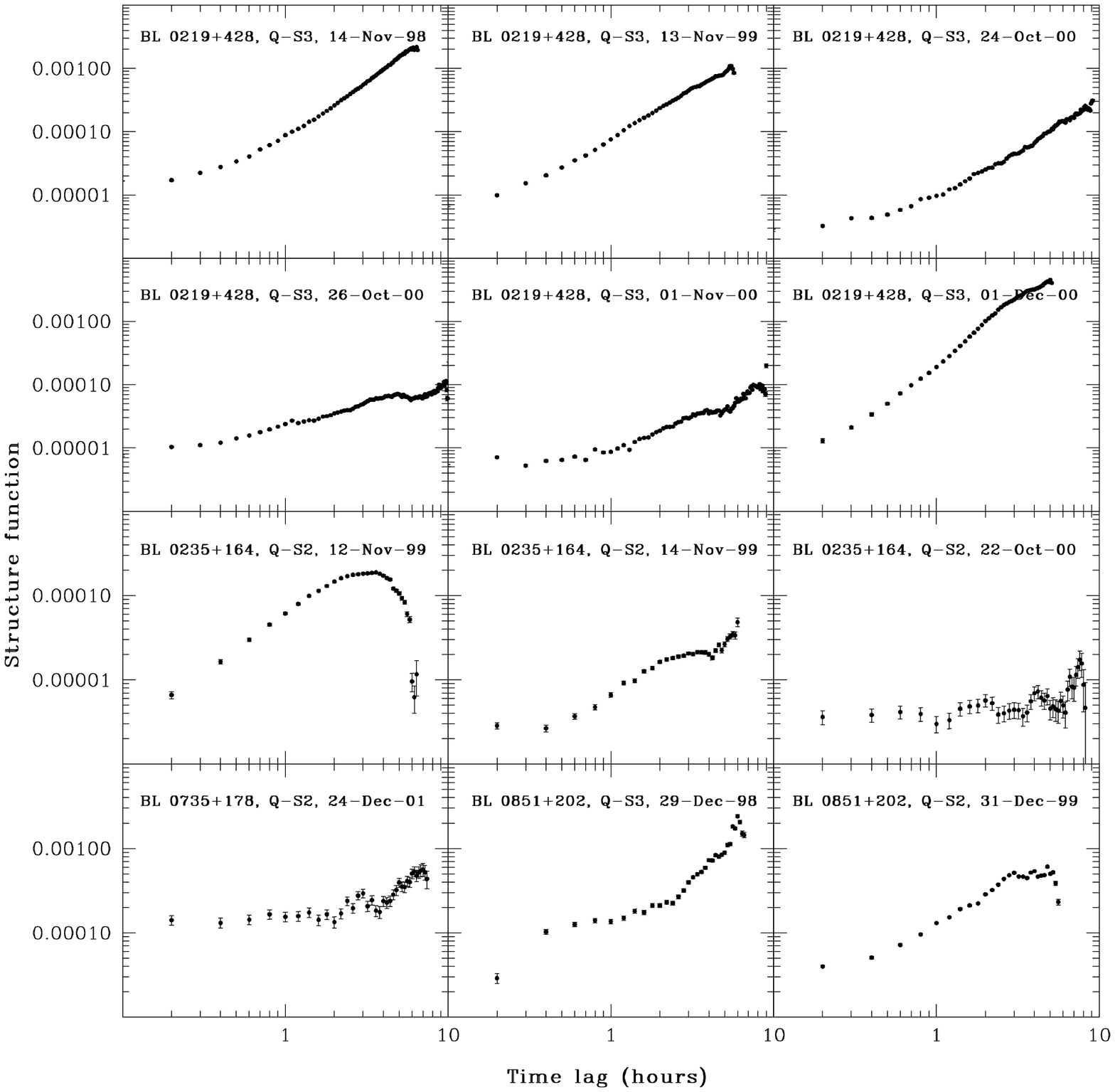,width=18cm,height=19cm}

\vspace*{-0.2cm}
\noindent {\bf Figure 2.} First order structure function of BL Lacs and CDQs which have shown
INOV on the marked epoch against the indicated comparison star. 
\end{figure*}

\begin{figure*}
\vspace*{-0.5cm}
\hspace*{-1.0cm}\psfig{file=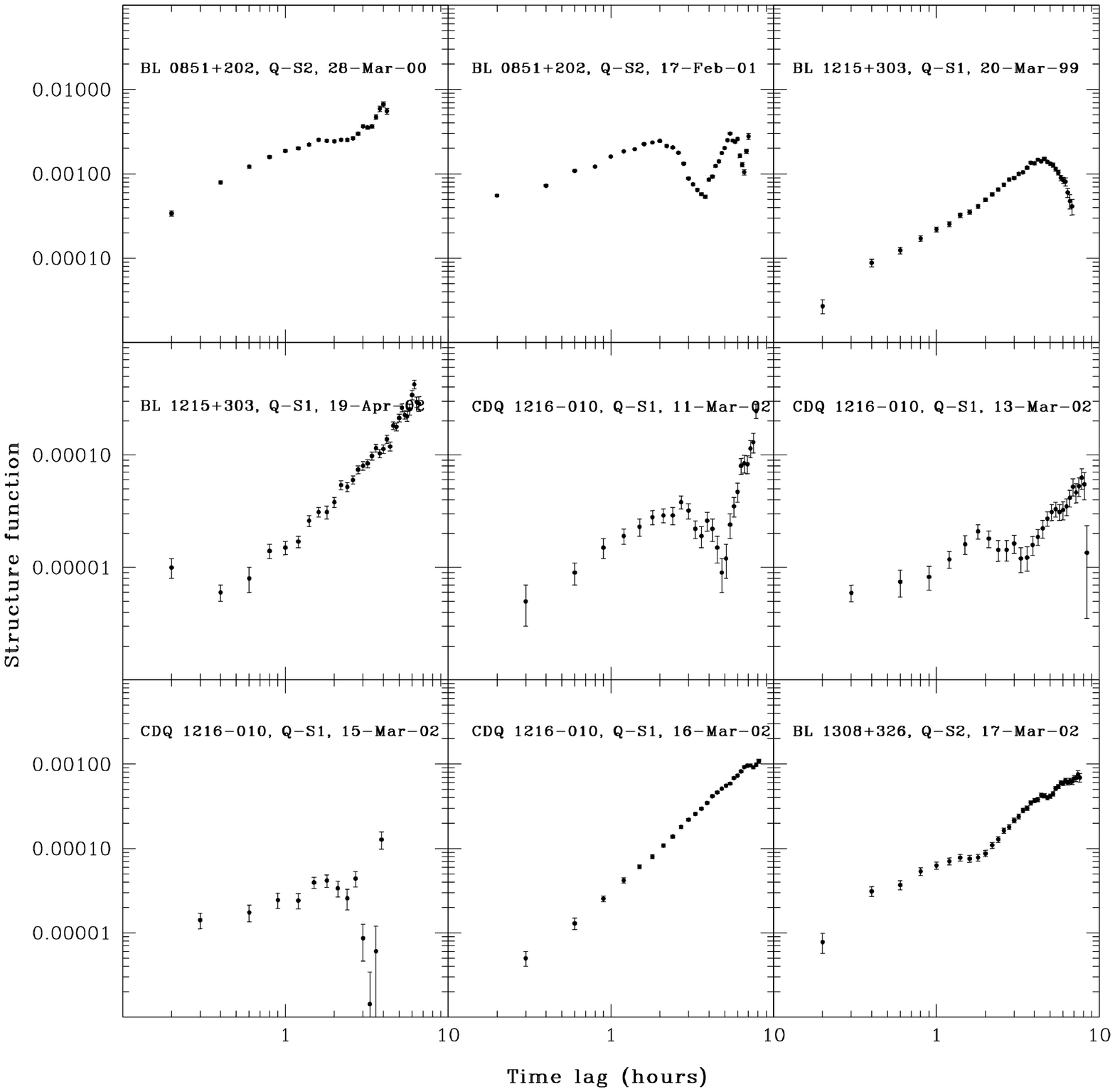,width=18cm,height=19cm}

\vspace*{-0.2cm}
\noindent {\bf Figure 2.} {\it Continued}
\end{figure*}

\noindent {\bf BL 0219+428}:  This is 
the best observed blazar in our sample, with seven
epochs of monitoring with durations between five and
ten hours, over the period from November 1998
to December 2000. Fig.\ 1 shows the DLCs for the six of 
these epochs when INOV was seen. On two of these 
epochs in October 2000, we also have I band
monitoring in addition to the default R band 
monitoring. On both of those nights the INOV in I and
R bands is found to be strongly correlated. The overall
INOV amplitudes on the six nights range between 
2 and 8\%. On 13 November 1999 and 1 December 2000, 
fairly abrupt changes in the slope of the DLCs were seen
during the 5 to 6 hours of continuous monitoring, and 
the INOV amplitudes recorded on these two nights
are also the largest observed for this object (Table 3). Another 
remarkable feature is that on 13 November 1999 at
19.6 UT a 1.5\% downward ``glitch" occurred within a 
time span of less than 10 minutes, followed by the onset
of steady fading. An expanded version of this part of 
the DLC is shown in the top panel of column 1 
in Fig.\ 1.  The SF plots for the six nights are 
smoother than usual, although  significant flattenings
are apparent on the nights of 01 November 2000 and 
26 October 2000, indicating time scales of around 4 hours (Fig.\ 2).
Both brightening and fading were noted in this BL Lac's LTOV.
The blazar faded by 0.70 mag during our first three epochs
of observations (between 14 November 1998 and 24 October 2000).
This was followed by a brightening of 0.34 mag within a week
that encompassed two additional nights of study.
The object had dimmed by 0.06 mag when observed 23 days
later on 24 November 2000 and was found to have brightened 
by 0.35 mag when observed for the last time on 1 December 2000.

\noindent {\bf BL 0235+164}:  This well known
BL Lac  object was monitored on five epochs between November 
1998 and October 2000. The DLCs of the first and last epochs 
are quite noisy, owing to the less sensitive CCD chip in the
former case and faintness of the object in the latter case.
On each of the remaining three DLCs, INOV is clearly seen
against all three comparison stars, with amplitudes ranging
between 7.6\% and 12.8\% (Fig.\ 1). The derived SFs for the two
nights in November 1999 have good S/N and in both cases
a flattening is noticed corresponding to time scale of
around 3 hours (Fig.\ 2). INOV monitoring of this 
object has also been recently reported by Xie et al.\ (2001). 
A particularly interesting feature of the DLCs on 
12 November 1999 is the rapid brightening of this object
by about 2\% within about 22 minutes (which corresponds to 
only about 11 minutes in the rest frame). 
This BL Lac also showed significant LTOV, with changes in both 
directions during the slightly under two year period during which 
we observed it. It had initially brightened by 0.67 mag between 
the first two epochs of observations (13 November 1998 and 12 
November 1999) followed by a dramatic 0.83 mag fading in two days
(12 November 1999 to 14 November 1999). Additional
fading by 0.45 mag was noticed when it was
observed a year later on 22 October 2000. This was 
again followed by a fading of 0.35 mag in the following
week (between 22 October 2000 and 28 October 2000). This blazar
had thus shown very large peak to peak LTOV 
during our observations: 1.63 mag within a year.

\noindent {\bf BL 0735+178}:  This BL Lac object was observed on 
four nights between December 1998 and December 2001. Although the 
data quality was generally good, on no occasion was a clear detection 
of INOV made, though there is marginal evidence for variations on 24 
December 2001 (Fig.\ 1). This behaviour is in clear contrast to that 
found here for the remaining five BL Lacs in our sample. Still, 
substantial LTOV was noted in this BL Lac.  It showed significant
brightening by 1.15 mag over the course of  our first three epochs of 
observations within two years. The blazar had faded by 0.43 mag from 
that peak over the next year when last monitored on 24 December 2001.

\noindent {\bf BL 0851+202}:  A recent detection of INOV of this well 
known BL Lac object has been reported by Ghosh et al.\ (2000). We 
monitored this object on four nights between December 1998 and February 2001 
and found it to show INOV on all four nights (Fig.\ 1). On each night 
features can be seen in the SF, corresponding to time scales between 1.5 and 
3.0 hours (Fig.\ 2). In addition, a ``periodicity'' of $\sim$ 3.8 hours is 
present in the SF for the night of 17 February 2001 (Fig.\ 2; Table 3); however, 
we note that this corresponds to only two peaks in the light curve (Fig.\ 1) 
and so need not be a genuine (quasi-)periodicity. The most remarkable feature 
on the DLC on 28 March 2000 is a $\sim$4\% jump in brightness against all 
three comparison stars within about an hour (at around 19 UT), which is also 
borne out in the SF analysis. This blazar also showed very large amplitude LTOV: 
it faded by 0.6 mag in a year between 29 December 1998 and 31 December 1999. 
However it then brightened during our remaining three epochs of observations.  
A total brightening of 1.63 mag was observed between 31 December 1999 and 17 Feb 2001.

\noindent {\bf CDQ 0955+326}: The only one of the three nights of monitoring this
quasar which showed possible indications of variability was 5 March 2000. Against 
all three comparison stars the quasar showed a steady decline of about 0.5\% over
6.9 hours (Fig.\ 1). Some LTOV of this quasar can be ascertained from our three 
epochs of observations which cover a span of about one year. The quasar brightened 
by 0.05 mag within a year between 19 February 1999 and 3 March 2000 and was found to 
remain at essentially the same level when observed two days later.

\noindent{\bf CDQ 1128+315}: No INOV was detected. Our three epochs of observations 
span about a year. Between 18 January 2001 and 9 March 2002 the quasar brightened 
by 0.14 mag; it remained at the same brightness level over the next 24 hours.

\noindent {\bf BL 1215+303}: This BL Lac object was monitored on five nights, of 
which only four nights were considered for INOV as the data on the remaining one 
night are of moderate quality. On 20 March 1999 the BL Lac showed clear evidence 
of variability with peak-to-peak amplitude of $\sim$ 4 to 5\% with a 2\% change 
noticed within $\sim$ 0.5 hour. On 19 April 2002 the DLCs showed a steady brightening 
of 2\% over the 6.8 hours of observations (Fig.\ 1). On 20 March 1999, the SF derived 
from 7.0 hours of observations shows a steady rise with a turnover corresponding to a 
time scale of 4.2 hours; no such SF turnover is present on 19 April 2002 (Fig.\ 2; Table 3).
The five epochs of monitoring observations of this BL Lac covered a temporal baseline 
of over three years (from 20 March 1999 to 19 April 2002). The quasar exhibited LTOV by 
fading progressively during the first four epochs, becoming fainter by 0.53 mag between 
20 March 1999 and 25 April 2001. By 19 April 2002 it had brightened again by 0.11 mag.

\noindent {\bf CDQ 1216$-$010}:  This highly polarized CDQ was monitored on four nights 
during March 2002 and clear INOV was detected on all the four nights against both 
comparison stars (Fig. 1). The SF for 16 March 2002 shows a linear rise all the way to 
the time lag of $\sim$8 hours. On each of the earlier three nights a ``periodicity'' 
signal of $\sim$ 3 to 4 hours is seen in the SF (Fig.\ 2; Table 3). This quasar was 
observed over a time baseline of only 6 days; nonetheless, the source is found to show 
significant inter-night variability. The quasar brightened by 0.02 mag between the 
first two epochs (11 March 2002 and 13 March 2002) followed by 0.11 mag fading
during the next two days. It once again had brightened (by 0.14 mag over the next 24 
hours) when first observed on 16 March 2002; it continued to brighten throughout that night.

\noindent {\bf CDQ 1225+317}: This most distant quasar in our sample did not show any 
evidence of INOV in our three roughly equally spaced epochs of observations over a span 
of two years (between 7 March 1999 and 20 April 2001). However, it showed LTOV, wherein 
a $\sim$2\% brightening was noticed in a year between April 2000 and April 2001. One might 
attribute this lack of detection to a combination of only three nights of observation
and the high redshift, which means that we were examining a
relatively short period in the rest frame of the CDQ.

\noindent {\bf BL 1308+326}:  This BL Lac object was monitored on six epochs between 
March 1999 and May 2002; however, it was sufficiently bright for the purpose of INOV 
detection on just the last three epochs, which fall in March, April and May 2002. 
INOV was detected only on the night of 17 March 2002 showing a gradual fading by 
about 2\% (Fig.\ 1), as well as possible very short-term variations. The SF
shows a plateau corresponding to a time scale of about 1.5 hours (Fig.\ 2). 
Large amplitude LTOV was seen. The quasar was relatively faint during the initial 
three epochs of our observations. A fading of 0.21 mag was found within the
first year (23 March 1999 and 26 April 2000) and it was found in the same level 
when observed a week later on 3 May 2000. However the quasar had brightened
by 1.9 mag when observed about 2 years later on 17 March 2002. It faded once again 
in about a month by 0.44 mag and then brightened by 0.03 mag over 12 days when 
observed for the last time on 2 May 2002.

\noindent {\bf CDQ 1309+355}: This quasar was observed for three epochs and showed 
no INOV, although it did exhibit LTOV. Between the first two epochs, separated by 
about two years, the quasar decreased in brightness by 0.10 mag, and remained at 
the same brightness level over the next 24 hours.

\subsection{Duty cycle (DC) of intra-night optical variability}

Duty cycles of INOV were calculated following the definition of Romero et al.\ (1999). 
Since most AGNs do not display variability on each night, duty cycles are best estimated 
not as a fraction of the variable objects found within a given class, but as the ratio
of the time over which objects of the class are seen to vary to the total observing time 
spent on monitoring the objects in the class. It is thus given as 
\begin{equation}
DC = 100 \frac{\sum_{i=1}^{n} N_i(1/\Delta t_i)}{\sum_{i=1}^{n}(1/\Delta t_i)}
\ \     \%,
\end{equation}
\noindent where $\Delta t_i = \Delta t_{i,obs}(1 + z)^{-1}$ is the duration (corrected 
for cosmological redshift) of an $i^{th}$ monitoring session of the source out of a 
total of $n$ sessions for the selected class, and $N_i$ equals 0 or 1, depending on 
whether the object was  non-variable or variable, respectively, during $\Delta t_i$.   
Results for the different classes (CDQ-LP: low polarization CDQs; CDQ-HP: high 
polarization CDQs; CDQ-ALL: CDQs; BL: BL Lacs) are shown as histograms in Fig.\ 3. The 
shaded portions refer to the subsets showing large variability ($\psi > 3$\%), while the
blank portions refer to the INOV detections of lower amplitudes; the cases of 
probable INOV are not included in these histograms.  
\begin{figure}
\vspace*{-0.7cm}
\hspace*{-0.5cm}\psfig{file=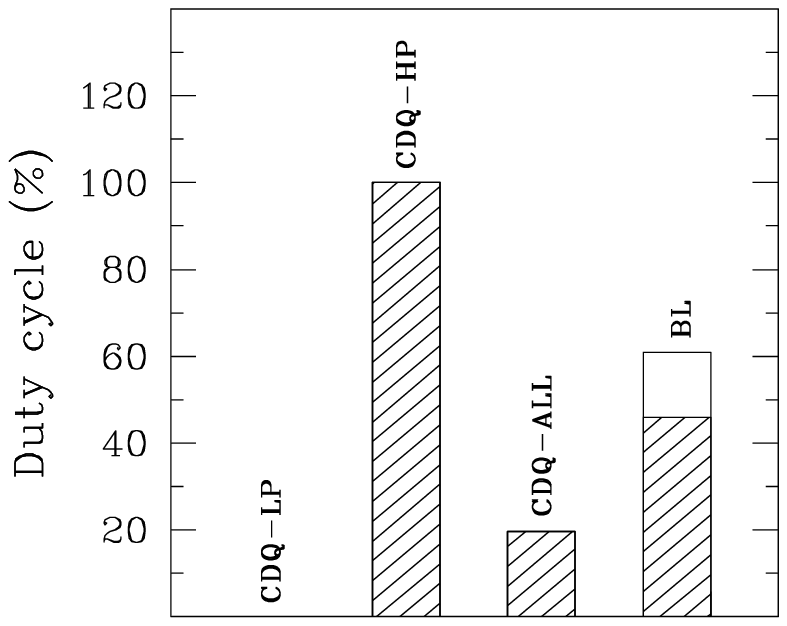}

\vspace*{-0.2cm}
\noindent{\bf Figure 3.}  Duty cycles for various classes of blazars and for
different amplitude ranges: shaded for $\psi > 3$\%; open for $\psi < 3$\%.
\end{figure}

The results presented here allow for the first time a comparison
of the DC of INOV for the three blazar classes, namely: BL Lacs; 
high-polarization CDQs, CDQ-HP; and low-polarization CDQs, CDQ-LP
(although our sample contains just one CDQ-HP, viz., 1216-010 (Table 1), 
a significant statement about it is possible, since it was monitored on four 
nights with an average duration of 7.1 hours per night).
In fact, the comparisons can now be made for ranges of variability 
amplitudes, which we take as $\psi$ $<$ 3\% and $\psi$ $>$ 3\%. In GSSW03 we
presented a similar comparison for RQQs and BL Lac objects
monitored in our program. We see that the DC for INOV detection is high both 
for BL Lacs (61\%) and for the CDQ-HP source (100\%), but much lower for the 
CDQ-LP sources. Thus, there appears to be a close link between the INOV and 
the polarized component of optical emission (which is commonly attributed to 
shocks in the relativistic non-thermal jets). While this pattern is highly 
suggestive, we must caution that a more definitive claim would require that 
the classification of the blazars into the two polarization classes be based 
on measurements made simultaneous to the optical flux monitoring and that 
the number of CDQ-HP sources be increased.

Further, it is interesting to note from Fig.\ 3 that not only do 
the BL Lacs and CDQ-HP exhibit high DC of INOV but also the probability 
of observing large amplitude INOV ($\psi >$ 3\%) is high (DC $\simeq 0.5$) and 
probably very similar to the probability of small amplitude INOV ($\psi <$ 3\%),
when we note that our lower limit for clear detections of INOV is $\psi \simeq $ 1\%. 

Any successful model of INOV should be able to explain this rather
unexpected behaviour found here for the blazars.

\section{Conclusions}

The observations reported here present the first systematic
study of the INOV characteristics of the two Doppler beamed AGN 
classes, CDQs and BL Lacs, although  studies of individual sources 
of these types have been made over the past decade (e.g., Noble et 
al.\ 1997).  We monitored both classes of blazars in our sample 
with equally high sensitivities and for comparable time durations;
the mean monitoring duration for BL Lacs was 6.5 hours over 31 
nights, while for CDQs it was 6.7 hours over 16 nights.

The major result of this observational programme is our finding that the
duty cycles of INOV detection for weakly polarized CDQs and BL Lacs 
are strikingly different. BL Lacs show a high DC of $\sim$60\%, in 
contrast to CDQs-LP for which the DC is found to be 0\% as none of them
showed INOV at the level of 1 \%. Including the one case of probable variable 
(0955+326 on 5 March 2000) raises this DC to 6\% only.
This  value is nominally lower than the DCs we have found for
RQQs and LDQs ($\sim$ 15\%; Stalin et al.\ 2003a; GSSW03). At the same time,
we find a close resemblance, both in amplitude and duty cycle of INOV,
between the one CDQ-HP in our sample and the BL Lacs. Thus it appears that the mere 
presence of a prominent (and hence presumably Doppler boosted) radio core does 
not guarantee INOV; instead, the more crucial factor appears to be
the optical polarization of the core emission. Such highly polarized emission 
is normally associated with shocks in a relativistic jet, so this may 
not be surprising. Of course the number of sources in our sample is not
large, and moreover the polarization measurements were made long before 
the intra-night optical monitoring reported here. Both these shortcomings 
need to be overcome in subsequent studies in order to place the present 
results on a firmer basis.

For most of the sources that do show INOV, some type of time scale of 
a few hours (observer's frame) is frequently detected.  Although, in a 
few cases, the structure function indicates the presence of several hour 
``periodicities'', we stress that these may well correspond to the 
time gap between multiple flares (or perturbations on bigger, slower 
flares). None of the light curves are long enough to actually detect 
putative real periodicities (or quasi-periodicities) of longer than
1 or 2 hours.  Probing such phenomena would require continuous 
monitoring through coordinated observations by many observatories,
well separated in longitude, or by a space observatory. 

\section*{Acknowledgments}
We thank the anonymous referee for suggestions that have significantly
improved the presentation of these results. This research has made use 
of the NASA/IPAC Extragalactic Database (NED), which is operated by 
the Jet Propulsion Laboratory, California Institute of Technology, under
contract with the National Aeronautics and Space Administration. CSS thanks 
NCRA for hospitality and use of its facilities. PJW is grateful for 
continuing hospitality at the Department of Astrophysical Sciences at Princeton University;
his efforts were partially supported by Research Program Enhancement funds at GSU.

\end{document}